\documentclass[aps, prl, reprint, superscriptaddress]{revtex4-1}

\usepackage{amsmath,amssymb}
\usepackage{xcolor}
\usepackage{hyperref,graphicx}

\newcommand{\ket}[1]{|#1\rangle}
\newcommand{\bra}[1]{\langle#1|}
\newcommand{\suppinf}{Supplemental Material}
\newcommand{\figref}[1]{Fig.~\ref{#1}}

\newcommand{\IFNCNR}{Istituto di Fotonica e Nanotecnologie, Consiglio Nazionale delle Ricerche, P.za Leonardo da Vinci 32, I-20133 Milano, Italy}
\newcommand{\UNIROMA}{Dipartimento di Fisica, Sapienza Universit\`{a} di Roma, P.le Aldo Moro 5, I-00185 Roma, Italy}
\newcommand{\POLIMI}{Dipartimento di Fisica, Politecnico di Milano, P.za Leonardo da Vinci 32, I-20133 Milano, Italy}

\newcounter{suppEq}
\newenvironment{suppEq}{\refstepcounter{suppEq}\equation}{\tag{S-\thesuppEq}\endequation}
\newenvironment{suppAlign}{\refstepcounter{suppEq}\align}{\tag{S-\thesuppEq}\endalign}

\begin{document}

\title{Fermionic statistics suppresses Fano resonances}

\author{Andrea Crespi}
\affiliation{\IFNCNR}
\author{Linda Sansoni}
\affiliation{\UNIROMA}
\author{Giuseppe Della~Valle}
\affiliation{\POLIMI}
\affiliation{\IFNCNR}
\author{Alessio Ciamei}
\affiliation{\UNIROMA}
\author{Roberta Ramponi}
\affiliation{\IFNCNR}
\affiliation{\POLIMI}
\author{Fabio Sciarrino}
\email{fabio.sciarrino@uniroma1.it}
\author{Paolo Mataloni}
\affiliation{\UNIROMA}
\author{Stefano Longhi}
\email{stefano.longhi@polimi.it}
\affiliation{\POLIMI}
\affiliation{\IFNCNR}
\author{Roberto Osellame}
\email{roberto.osellame@ifn.cnr.it}
\affiliation{\IFNCNR}
\affiliation{\POLIMI}


\begin{abstract}
Fano resonances and bound states with energy in the continuum are ubiquitous phenomena in different areas of physics. 
Observations, however, have been limited so far to single-particle processes. In this work we experimentally investigate the multi-particle case and observe Fano interference in a non-interacting two-particle Fano-Anderson model by considering propagation of two-photon states in engineered photonic lattices. 
We demonstrate that the quantum statistics of the particles, either bosonic or fermionic,  strongly affects the decay process. Remarkably, we find that the Fano resonance, when two discrete levels are coupled to a continuum, is suppressed in the fermionic case.
\end{abstract}

\maketitle

Decay of excited states has been a topic of great interest since the early times of quantum mechanics \cite{fonda1978rpp}. An excited state is usually modelled by a Breit-Wigner resonance, which is the universal hallmark of unstable states. However, quantum interference, arising from wave function superposition, can lead to different manifestations of resonant behavior for unstable states when multiple paths are possible.  This is the case of the Fano resonance \cite{fano1961pr,fano1986}.

Fano's model\cite{fano1961pr,fano1986} is a landmark in modern physics. Firstly developed to explain the behaviour of electrons scattered by excited atoms \cite{fano1961pr}, it was later adopted to explain phenomena in a number of different physical systems \cite{miroshnichenko2010rmp}, such as ultracold gases and Bose-Einstein condensates (where the Fano resonance is usually referred to as Feshbach resonance \cite{chin2010rmp}), semiconductors, quantum dots and mesoscopic systems \cite{yoon2012prx,yoon2012b,yoon2012c}, and plasmonic nanostructures \cite{lukyanchuk2010nmat}.  Fano interference is observed when different decay channels interfere, giving rise to broadening and asymmetric deformations of natural line shapes. In general, the destructive interference between different decay channels is associated to the formation of bound states in the continuum \cite{fano1961pr,sudarshan1988}, which inhibit the complete decay of the excited state. The interplay between bound states in the continuum and Fano/Feshbach resonances has been highlighted in several works (see, for instance, \cite{FB0,FB1,FB2,FB3}). 
Experimental studies on Fano resonances \cite{miroshnichenko2010rmp}, quantum decay processes \cite{willkinson1997nat,fischer2001prl} and bound states in the continuum \cite{capasso,deo,kepalle1,weimann,corrielliPRL2013} have focused up to now on single-particle dynamics. Interestingly, recent works \cite{taniguchi2011pre,delcampo2011pra,garciacalderon2011pra,rontani2012prl,pons2012pra,longhi2012pra,zurn2013prl} showed that particle statistics and contact interactions can deeply modify the decay dynamics. Fermions and bosons may show very different decay behaviour, in particular in many cases fermions tend to decay faster \cite{taniguchi2011pre,delcampo2011pra,garciacalderon2011pra}. However, no experimental observation of this phenomenon has been reported yet.

In this work we investigate, experimentally, the decay process of two non-interacting particles to a common continuum, by probing an engineered photonic lattice with two-photon states. The lattice, consisting of a three-dimensional array of coupled optical waveguides, is fabricated in a glass substrate by femtosecond laser micromachining \cite{fslaserwriting,fslaserwriting_b,fslaserwriting_c}. While the bosonic dynamics is naturally observed for identically polarised photons, an antisymmetric polarisation-entangled state of the two photons is used to simulate the fermionic behavior \cite{omar2006qw,lahini2010,sansoni2012prl,matthews2013scirep,cres13npo}. 

\begin{figure}[t]
\centering
\includegraphics[width=8cm]{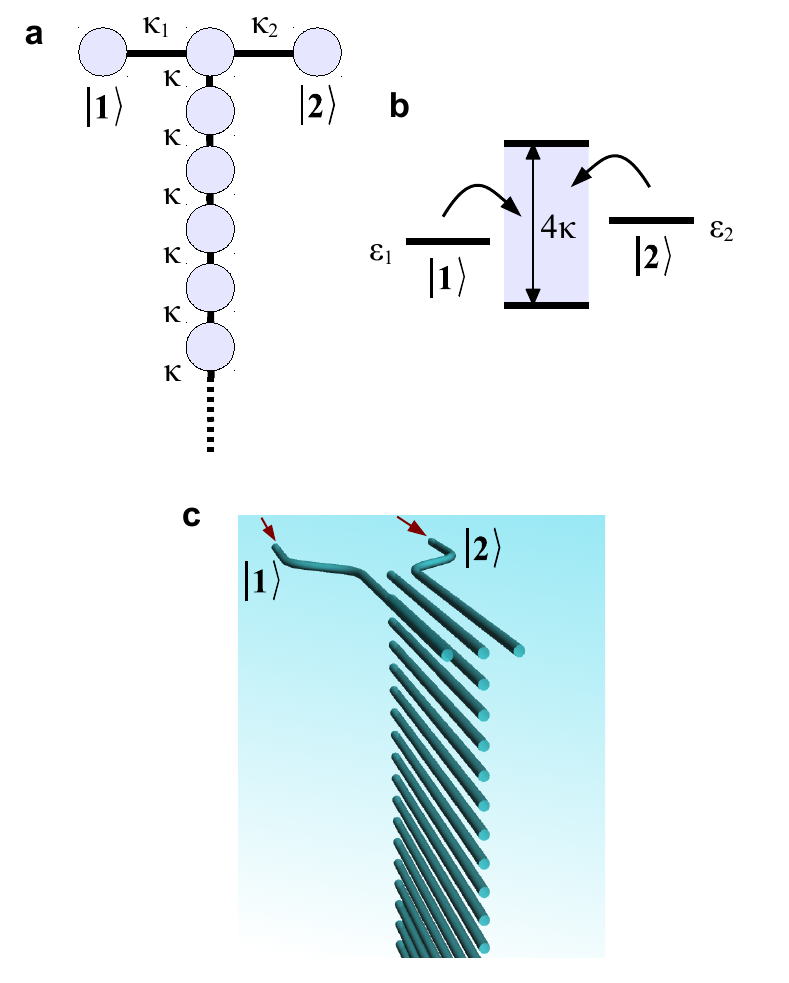}
\caption{Schematic of (a) two quantum wells side-coupled to a tight binding quantum wire, and (b) representation of the energy levels. (c) 3D rendering of the actual photonic structure employed in the experiments; two waveguides coupled to a vertical linear array represent the two discrete states coupled to the continuum. The two-photon state is launched in the photonic structure as indicated by the red arrows.} 
\label{fig:model}
\end{figure}

We focus on systems described by the Fano-Anderson \cite{fano1961pr,miroshnichenko2010rmp} or Friedrichs-Lee Hamiltonian \cite{friedrichs,lee}, which is  a paradigmatic model to study quantum mechanical decay, Fano interference phenomena and bound states in the continuum \cite{sudarshan1988,FB2,FB3,longhi2012pra}. The simplest case is provided by two discrete states coupled to a common tight-binding continuum of modes, i.e. a quantum wire \cite{longhi2012pra}. In detail, we consider a system composed of two sites $\ket{1}$ and $\ket{2}$, respectively with energy $\epsilon_1$ and $\epsilon_2$, side-coupled with hopping rates $\kappa_1$ and $\kappa_2$, to a common semi-infinite chain of coupled sites (a quantum wire), each with energy $\epsilon = 0$ (see \figref{fig:model}a-b). Thus, states $\ket{1}$ and $\ket{2}$ can decay by tunnelling to the common continuum given by the tight-binding lattice band of the quantum wire. The energy of the band spans the interval $-2 \kappa < E < 2 \kappa$, being $\kappa$ the hopping rate between two adjacent sites of the wire. In the following, $\epsilon_1$ and $\epsilon_2$ will be considered embedded into the continuum, i.e. $|\epsilon_{1,2}| < 2 \kappa$, and the coupling of the sites $\ket{1}$ and $\ket{2}$ to the continuum will be assumed as weak, i.e. $\kappa_{1,2} < \kappa$.

In our experiment, the sites correspond to optical modes of different waveguides, which interact through their evanescent field. The system is thus formed by two waveguides coupled to a linear waveguide array, according to the geometry of \figref{fig:model}c. The coupling coefficients $\kappa$, $\kappa_1$, $\kappa_2$ depend on the distance between the waveguides and can be tailored by carefully dimensioning the structure. The energies $\epsilon_1$, $\epsilon_2$ and $\epsilon$ correspond to the propagation constants of the waveguides themselves and can be tuned by varying the refractive index change in the different waveguides (for our direct laser writing process, this is achieved by tailoring the writing speed). In this photonic implementation, the temporal dynamics of the system is mapped onto the propagation distance $z$.
\begin{figure}[t]
\centering
\includegraphics[width=8cm]{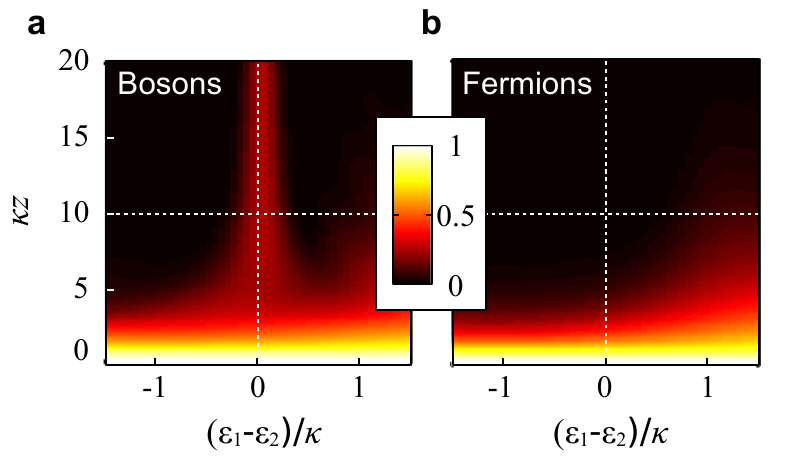}
\caption{Numerical simulations of the survival probability $P_S$ for different propagation lengths and energy detuning are reported for the case of (a) bosonic or (b) fermionic particles, respectively. Propagation and energy coordinates are normalized with respect to $\kappa$, and the parameters used for the simulations are $\kappa_1=\kappa_2=0.4 \kappa$ and $\epsilon_1=\kappa$.}
\label{fig:simulation}
\end{figure}

The evolution of the particle-creation operators $a^{\dagger}_j$ for the various modes is governed by the coupled-mode equations:
\begin{align}
i \frac{d a^{\dagger}_{1,2}}{d z} &= \epsilon_{1,2} a^{\dagger}_{1,2} + \kappa_{1,2} a^{\dagger}_3 \nonumber\\
i \frac{d a^{\dagger}_3}{d z} &= \kappa_1 a^{\dagger}_1 + \kappa_2 a^{\dagger}_2 + \kappa a^{\dagger}_4 \nonumber \\
i \frac{d a^{\dagger}_j}{d z} &= \kappa a^{\dagger}_{j-1} + \kappa a^{\dagger}_{j+1} \qquad j \geq  4  \label{eq:coupledMode}
\end{align}
where $a^{\dagger}_1$ and $a^{\dagger}_2$ refer to the modes $\ket{1}$ and $\ket{2}$, while $a^{\dagger}_j$ with $j>2$ refer to the modes of the linear array. The equivalence between the semi-infinite lattice model, described by the operator equations \eqref{eq:coupledMode}, and the Fano-Anderson model can be readily established by an operator transformation from the Wannier to the Bloch basis representation \cite{FB2,FB3,longhi2012pra}. An interesting property of this system is the existence of one bound state in the continuum when $\epsilon_1=\epsilon_2$. In fact, it is easy to observe that in this case the operator $b^{\dagger} = e^{i \epsilon_1 z} \left(a^{\dagger}_1/\kappa_1 - a^{\dagger}_2/\kappa_2\right)$ satisfies ${d b^{\dagger}}/{dz} = 0$,
i.e. population of the dressed state described by $b^\dagger$ does not decay. In the Bloch basis of operators, the bound state can be interpreted as a result of a destructive Fano interference between different decay channels \cite{FB2,FB3,longhi2012pra}. This leads to fractional decay when a single particle (photon) is placed in either site (waveguide) $\ket{1}$ or $\ket{2}$. 
We are studying here the two-particles case, considering an initial state $\ket{\Psi(0)}$ of the system excited with one photon in $\ket{1}$ and one photon in $\ket{2}$. Quantum decay is described by the survival probability:
\begin{equation}
P_S (z) = | \langle \Psi(0)| \Psi(z) \rangle |^2
\label{eq:ps}
\end{equation}
which is the probability that, at a propagation distance $z$, both particles are still in the initial state and none has decayed into the continuum. From an experimental point of view, $P_S$ corresponds to the probability of coincidence detection of two photons in output modes $\ket{1}$ and $\ket{2}$.
\begin{figure*}[t!]
\centering
\includegraphics[width=15cm]{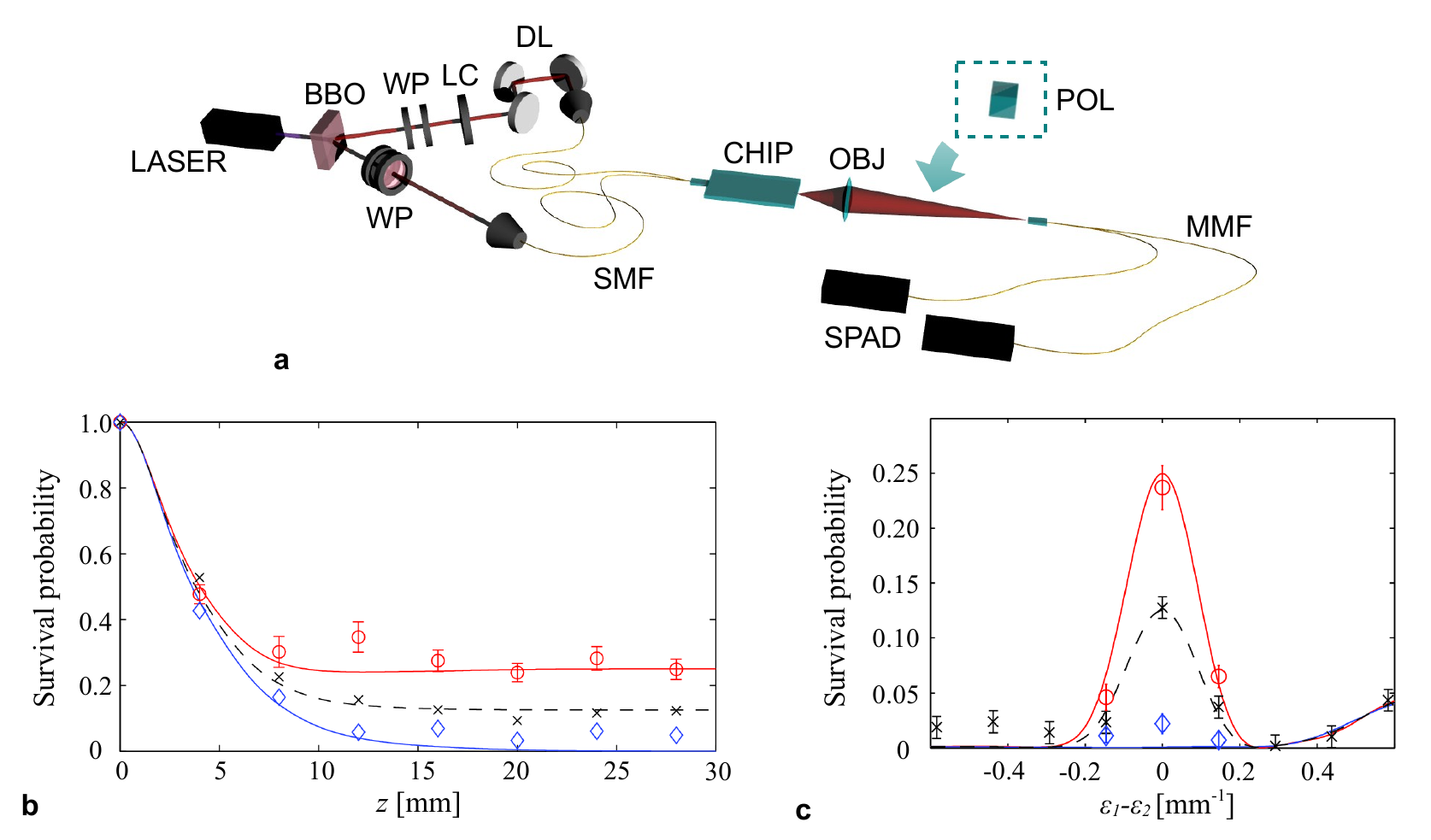}
\caption{(a) Experimental setup for two-photon measurements: two-photon entangled states are generated by spontaneous-parametric down-conversion in a BBO crystal. The generated state is tuned by waveplates (WP) and a liquid crystal retarder (LC); photons are coupled to single-mode fibers (SMF) and injected into the chip. A microscope 5$\times$ objective (OBJ) collects the output light and images the two lateral modes directly onto the entrance facets of two multi-mode fibers (MMF) in a 250~$\mu$m pitched fibre array. The fibres are connected to single-photon avalanche photodiodes (SPAD) for detection. A delay line (DL) controls the temporal indistinguishability of the photons. Measurements for $\ket{VV}$ state are performed by injecting a $\ket{HH}+\ket{VV}$ state in the chip and post-selecting with a polariser (POL). (b-c) Experimental maps of the survival probability for two bosons (red circles), two fermions (blue diamonds), two distinguishable particles (black crosses). Curves of numerical simulation are also shown. In (b) $\epsilon_1=\epsilon_0$ and $z$ is varied, which corresponds to the vertical dashed lines in the maps of \figref{fig:simulation}a-b. In (c) $z = 20$~mm is kept fixed while $\epsilon_2$ is varied, which corresponds to the horizontal dashed lines in the maps of \figref{fig:simulation}a-b.
Where not shown, errorbars are smaller than marker size.}
\label{fig:setup}
\end{figure*}

By exciting the system with two identically polarised photons the natural bosonic behaviour is observed. On the  contrary, if the system is excited with an antisymmetric polarisation-entangled two-photon state, $P_S (z)$ has the same expression as if we were injecting two identical particles with fermionic statistics, i.e. the operators $a^{\dagger}_j$ satisfy fermionic commutation rules (see e.g. Refs. \onlinecite{sansoni2012prl,matthews2013scirep} and the specific discussion in the \suppinf). 
Figure \ref{fig:simulation}a-b show numerical simulations of survival probability $P_S (z)$ for a system described by \eqref{eq:coupledMode}, in normalized coordinates, for initial two-particle states either bosonic or fermionic. A striking difference between the bosonic and fermionic behaviour is evident, as will be experimentally demonstrated in the following.

As a first experiment, we concentrate on the investigation of the quantum decay in the case $\epsilon_1=\epsilon_2$. We have fabricated several structures as that reported in \figref{fig:model}c, yielding $\kappa_1=\kappa_2=0.2$~mm$^{-1}$, $\kappa = 0.5$~mm$^{-1}$, $\epsilon_1=\epsilon_2=0.5$~mm$^{-1}$, and different lengths $z$ of the array. In our realization the linear array is composed of 25 waveguides. Details on the fabrication process by femtosecond laser direct writing are given in the \suppinf. Each structure allows us to photograph the evolution at a specific propagation distance $z$.

The system is first characterized by launching laser light separately in waveguides $\ket{1}$ and $\ket{2}$, and imaging the output facet with a CMOS camera. The fraction of light remaining in the launch waveguides is measured. This corresponds to investigating the single-particle behaviour, when the particle is initially on mode $\ket{1}$ or $\ket{2}$. Survival probability $P_{S,clas}$ for two classical, distinguishable particles is easily calculated from the product of two single-particle experimental distributions (since the two particles are uncorrelated).

To experimentally characterize the system behaviour for two correlated particles, two photons at 810~nm wavelength, generated by a spontaneous parametric down-conversion source, are coupled to single-mode optical fibres and injected simultaneously in waveguides $\ket{1}$ and $\ket{2}$. Output light from the same waveguides is collected by an objective, coupled to multimode fibres and detected by single-photon avalanche photodiodes (see \figref{fig:setup}a).  Coincidence-detection counts, in equal temporal gates, are performed for different input states: indistinguishable vertically polarised photons, polarisation-entangled photons in antisymmetric state and distinguishable photons (the latter being generated by introducing, for each of the previous states, a temporal delay for one of the photons). These conditions correspond to identical bosons, identical fermions and distinguishable particles. The survival probability for identical bosons $P_{S,bos}$ and fermions $P_{S,fer}$ is then retrieved from experimental measurements as follows:
\begin{align}
P_{S,bos} &= \frac{C_{VV}}{C_{VV,dist}} P_{S,clas} \\
P_{S,fer} &= \frac{C_{ent}}{C_{VH,dist}} P_{S,clas}
\end{align}
where $C_{VV}$ are the coincidence counts for the $\ket{VV}$ state, $C_{VV,dist}$ the corresponding counts when one photon is delayed, $C_{ent}$ are the coincidence counts for the entangled  $\left( \ket{HV} - \ket{VH} \right) / \sqrt{2}$ state, $C_{VH,dist}$ the corresponding counts when one photon is delayed.

Figure \ref{fig:setup}b reports the experimentally characterized survival probabilities for the different $z$ and different input states, compared to numerical simulations for the same system. While $P_{S,bos}$ shows a fractional decay owing to the existence of a bound state, $P_{S,fer}$ shows a full decay. The experimental points for fermions do not reach exactly zero due to imperfections in preparing the entangled state and slight polarisation dependence of the photonic device, which introduces photon distinguishability. However, such results clearly show that Fano interference, responsible for the existence of a bound state and fractional decay, is suppressed for two fermions. This is a signature of the Pauli exclusion principle 
and can be explained by observing that no more than one fermion can be accommodated into the (single) dressed bound state, while the other fermion necessarily decays into the state continuum.

To better highlight the difference between bosonic and fermionic behaviour in the decay process, Fano-like profiles are measured from the survival probability
as the detuning of the energy levels of the two discrete states is varied \cite{FB2,FB3}. We fabricated and characterized other photonic structures, with fixed length $z = 20$~mm, the same $\kappa_1$, $\kappa_2$, $\kappa$ as in the previous experiments, $\epsilon_1 = 0.5$~mm$^{-1}$ and different values for $\epsilon_2$. Results are shown in \figref{fig:setup}c. For bosonic particles, the survival probability shows a clear peak at $\epsilon_1=\epsilon_2$. This confirms that the fractional decay observed in the previous experiment is indeed due to a Fano resonance condition, which vanishes for detuned energies of the two discrete levels. Interestingly this Fano resonance peak is absent when the two particles possess fermionic statistics.

It should be pointed out that a generalization of this model to more than two discrete levels coupled to the continuum, will result in a possibly higher number of bound states and the influence of the particle statistics may be more articulated. However, as long as the number of bound states in the continuum is lower than the number of particles, a fermionic statistics will always suppress the Fano resonance \cite{longhi2012pra}.

In conclusion, this work has presented an experimental study on the quantum decay process of two identical particles, initially on discrete states, into a common continuum. A profound difference in the bosonic and fermionic evolution is evidenced; in particular, a suppression of a Fano resonance condition is observed for the fermionic case. The capability to simulate multi-particle dynamics in discrete systems coupled to a continuum may enable the investigation of other multi-particle decay phenomena, such as multi-particle Zeno effects \cite{wang2008pra} and non-Markovianity \cite{lofranco2013ijmpb}.

\begin{acknowledgments}
This work was supported by ERC (European Research Council) Starting Grant 3D-QUEST (3D-Quantum Integrated Optical Simulation; grant agreement no. 307783 - www.3dquest.eu) and by the European Union through the project FP7-ICT-2011-9-600838 (QWAD — Quantum Waveguides Application and Development - www.qwad-project.eu).
\end{acknowledgments}

\cleardoublepage




\section*{\suppinf}

\subsection{Femtosecond laser waveguide writing}

Waveguides are fabricated in Eagle2000 (Corning) glass substrate. Pulses of 220~nJ energy and $\sim300$~fs duration, at 1~MHz repetition rate, from a Yb:KYW cavity dumped oscillator ($\lambda = 1030$~nm) have been employed, focused by a 0.6~NA microscope objective. Translation speed is varied depending on the desired propagation constant. In particular, in order to provide the different values of $\epsilon_2$ in the experiment of \figref{fig:setup}b of the Main Text, it was varied between 37~mm/s and 56~mm/s. A speed gradient was also employed in fabricating the vertical array, in order to compensate for the effect of spherical aberration at different depths and obtain, as a result, uniform propagation constants. The distance between waveguide $\ket{1}$ or $\ket{2}$ and the first waveguide of the array is 12~$\mu$m, the distance between adjacent waveguides of the vertical array is 10~$\mu$m. The shallower waveguides are buried 170~$\mu$m below the sample surface.

\subsection{Survival probability and particle statistics}

We consider the system represented in \figref{fig:model} of the Main Text, whose evolution is described by
\begin{suppAlign}
i \frac{d a^{\dagger}_{1,2}}{d z} &= \epsilon_{1,2} a^{\dagger}_{1,2} + \kappa_{1,2} a^{\dagger}_3 \nonumber\\
i \frac{d a^{\dagger}_3}{d z} &= \kappa_1 a^{\dagger}_1 + \kappa_2 a^{\dagger}_2 + \kappa a^{\dagger}_4 \nonumber \\
i \frac{d a^{\dagger}_j}{d z} &= \kappa a^{\dagger}_{j-1} + \kappa a^{\dagger}_{j+1} \qquad j \geq  4  \label{eq:coupledModeSI}
\end{suppAlign}
and an initial states with two particles, one in $\ket{1}$ and one in $\ket{2}$ respectively. Then:
\begin{suppEq}
\ket{\Psi (0)} = a^{\dagger}_1 (0) a^{\dagger}_2 (0) \ket{0}
\label{eq:inState}
\end{suppEq}
and, from the definition of survival probability
\begin{suppEq}
P_S (z) = | \langle \Psi(0)| \Psi(z) \rangle |^2
\label{eq:psSupp}
\end{suppEq}
one has:
\begin{suppEq}
P_S (z) = | \bra{0} a_2 (0) a_1 (0) a^{\dagger}_1 (z) a^{\dagger}_2 (z) \ket{0} |^2
\label{eq:psExpanded}
\end{suppEq}
By solving the system \eqref{eq:coupledModeSI} for a given $z$, one can calculate the elements of the scattering matrix $\mathcal{S} (z) = S_{n,j} (z)$ so that
\begin{suppEq}
  a^{\dag}_j (z) = \sum_{n=1}^{\infty} S_{n,j} (z) a^{\dag}_j
\end{suppEq}
Then, depending on the bosonic or fermionic commutation relations of the $a^{\dagger}_j$ operators, one finds:
\begin{suppEq}
P_{S,bos}(z)= |S_{1,1} S_{2,2}+S_{1,2}S_{2,1} |^2= |{\rm perm} \; \mathcal{S}(t)|^2
\label{eq:psboson}
\end{suppEq}
 for bosonic particles, or
\begin{suppEq}
P_{S,ferm}(z)= |S_{1,1}S_{2,2}-S_{1,2}S_{2,1} |^2= |{\rm det} \; \mathcal{S}(t)|^2
\label{eq:psfermions}
\end{suppEq}
for fermions.
On the other hand, for distinguishable particles one has:
\begin{suppEq}
P_{S,clas}(z)=\left| S_{1,1}S_{2,2} \right|^2+\left| S_{1,2} S_{2,1} \right|^2.
\end{suppEq}  

Considering now more specifically our optical experimental system, if we inject two identically polarised photons (an initial state that can be described by Eq. \eqref{eq:inState}), we are naturally dealing with the quantity given by Eq. \eqref{eq:psboson}, given the bosonic statistics of photons.

However, if we inject a polarisation entangled antisymmetric state:
 \begin{suppEq}
 \ket{\Psi(0)} = \frac{1}{\sqrt{2}} \left( a_{1,H}^{\dag} a_{2,V}^{\dag} -  a_{1,V}^{\dag} a_{2,H}^{\dag} \right) | 0 \rangle
\label{eq:inEntState}
 \end{suppEq}
the evolved state at a coordinate $z$ becomes:
  \begin{suppAlign}
  \ket{ \Psi(z) }  = &\frac{1}{\sqrt{2}} \left[  \left( S_{1,1} S_{1,2}- S_{1,1} S_{1,2} \right) {a}^{\dag}_{1,H} {a}^{\dag}_{1,V}  \right. \nonumber \\
 + &\left( S_{1,1} S_{2,2} - S_{2,1} S_{1,2} \right) {a}^{\dag}_{1,H} {a}^{\dag}_{2,V} \nonumber \\
 + &\left( S_{2,1} S_{1,2} - S_{1,1} S_{2,2} \right) {a}^{\dag}_{2,H} {a}^{\dag}_{1,V} \nonumber \\
+ &\left( S_{2,1} S_{2,2} -S_{2,1} S_{2,2} \right) {a}^{\dag}_{2,H} {a}^{\dag}_{2,V} \nonumber \\
 + &\left. ... \right] \ket{0}
\label{eq:outEntState}
  \end{suppAlign}
where we have assumed a polarisation independent scattering matrix $\mathcal{S}$ and where the dots indicate terms  containing creation operators $a^{\dag}_{j,H/V}$ with $ j \geq 3$ .
Then, if the survival probability \eqref{eq:psSupp} is calculated by using the expressions \eqref{eq:inEntState} and \eqref{eq:outEntState}, one finds that $P_S$ has exactly the form of \eqref{eq:psfermions} and a fermionic behavior is simulated.

%

\end{document}